\documentclass[12pt]{article}

\usepackage{geometry}                
\usepackage{amsmath,amssymb}
\usepackage{graphicx}
\usepackage[pdftex,colorlinks=false,urlcolor=blue,linkcolor=blue]{hyperref}

\usepackage{epstopdf}
\def\sigsi{\sigma_{\rm SI}}
\def\mz{m_Z}
\def\gev{~{\rm GeV}}
\def\tev{~{\rm TeV}}
\def\beq{\begin{equation}}

\def\bea{\begin{eqnarray}}
\def\eea{\end{eqnarray}}
\def\gam{\gamma}
\def\kap{\kappa}
\def\lam{\lambda}
\def\akap{A_\kap}
\def\alam{A_\lam}
\def\mhalf{m_{1/2}}
\def\mhu{m_{H_u}}
\def\mhd{m_{H_d}}

\def\omghsq{\Omega h^2}
\def\amu{a_\mu}
\def\damu{\delta \amu}
\def\br{{\rm BR}}

\def\hsm{h_{\rm SM}}

\def\rhi{R^{h_1}}
\def\rhii{R^{h_2}}
\def\to{\rightarrow}
\def\hi{h_1}
\def\hii{h_2}
\def\ai{a_1}

\def\mhi{m_{\hi}}
\def\mhii{m_{\hii}}
\def\mai{m_{\ai}}

\def\mh{m_h}

\def\stau{\widetilde \tau}
\def\cnone{\widetilde \chi_1^0}
\def\cpone{\widetilde \chi_1^\pm}
\def\mcnone{m_{\cnone}}
\def\mcpone{m_{\cpone}}
\def\hsm{h_{\rm SM}}

\def\msq{m_{\tilde q}}
\def\mgl{m_{\tilde g}}
\def\stopi{\tilde t_1}
\def\mstopi{m_{\stopi}}

\def\tanb{\tan\beta}

\def\lsim{\mathrel{\raise.3ex\hbox{$<$\kern-.75em\lower1ex\hbox{$\sim$}}}}
\def\gsim{\mathrel{\raise.3ex\hbox{$>$\kern-.75em\lower1ex\hbox{$\sim$}}}}

\def\bit{\begin{itemize}}
\def\eit{\end{itemize}}
\def\bec{\begin{center}}
\def\eec{\end{center}}

\begin{document}

\vspace*{6mm}

\begin{center}

{\LARGE The constrained NMSSM and Higgs near 125 GeV}\\[10mm]

{\large {John F.~Gunion}$^a$, {Yun~Jiang}$^a$, {Sabine~Kraml}$^b$}\\[6mm]

{\it
$^a$\,{Department of Physics, University of California, Davis, CA 95616, USA}\\
$^b$\,{Laboratoire de Physique Subatomique et de Cosmologie, UJF Grenoble 1,\\ 
CNRS/IN2P3, INPG, 53 Avenue des Martyrs, F-38026 Grenoble, France}
}

\end{center}

\hrule
\begin{abstract}
We assess the extent to which various constrained versions of the NMSSM are able to describe the recent hints of a Higgs signal at the LHC corresponding to a Higgs mass in the range $123-128\gev$.
\end{abstract}
\hrule

\vspace*{6mm}


The Large Hadron Collider (LHC) data from the ATLAS~\cite{atlashiggs}
and CMS~\cite{cmshiggs} collaborations suggests the possibility of a
fairly Standard Model (SM) like Higgs boson with mass of order
$123-128\gev$. In particular, promising hints appear of a narrow
excess over background in the $\gam\gam$ and $ZZ\to 4\ell$ final
states with strong supporting evidence from the $WW\to\ell\nu\ell\nu$
mode. While the ATLAS and CMS results suggest that the $\gam\gam$ rate
may be somewhat enhanced with respect to the SM expectation, this is
by at most one standard-deviation  ($1\sigma$).

In this Letter, we explore the ability or lack thereof of three
constrained versions of the next-to-minimal supersymmetric standard
model (NMSSM) to describe these observations while remaining
consistent with all relevant constraints, including those from LEP and
TEVATRON searches, $B$-physics, the muon anomalous magnetic moment,
$\amu\equiv (g-2)_\mu/2$, and the relic density of dark matter, $\omghsq$.

The possibility of describing the LHC observations in the
context of the MSSM has been explored in numerous papers, including~\cite{Baer:2011ab}\cite{Arbey:2011ab}\cite{Arbey:2011aa}\cite{Carena:2011aa}\cite{Buchmueller:2011ab}\cite{Akula:2011aa}\cite{Kadastik:2011aa}\cite{Cao:2011sn}\cite{Arvanitaki:2011ck}.
A general conclusion seems to be that if all the constraints noted
above, including $\amu$ and $\omghsq$, are imposed rigorously, then the MSSM---especially a constrained version such as the CMSSM---is hard pressed to yield a fairly SM-like light Higgs boson at $125\gev$. 
This is somewhat alleviated when 
the $\amu$ constraint is dropped~\cite{Baer:2011ab}\cite{Buchmueller:2011ab}. 
Overall, however, large mixing and large SUSY masses are needed to achieve $m_h\sim 125\gev$.  
There has also been some exploration in the context of the NMSSM~\cite{Hall:2011aa}\cite{Ellwanger:2011aa}\cite{Arvanitaki:2011ck}, showing that for completely general parameters there is less tension between a light Higgs with
mass $\sim 125\gev$ and a lighter SUSY mass spectrum. The study presented here will be done in the context of several constrained versions of the NMSSM with universal or semi-universal GUT scale boundary conditions. Results for a very
constrained version of the NMSSM, termed the cNMSSM, appear in \cite{Djouadi:2008yj}\cite{Djouadi:2008uj}\cite{Arbey:2011ab} --- we discuss comparisons later in the paper.

The three models which we discuss here are defined in terms of grand-unification (GUT) scale parameters as follows: 
{\bf I})~a version of the constrained NMSSM (CNMSSM) in which we adopt universal $m_0$, $\mhalf$, $A_0=A_{t,b,\tau}$ values but require  $\alam=\akap=0$, as motivated by the $U(1)_R$ symmetry limit of the NMSSM; 
{\bf II})~the non-universal Higgs mass (NUHM) relaxation of model {\bf I} in which $\mhu$ and $\mhd$ are chosen independently of $m_0$, but still with $\alam=\akap=0$; and 
{\bf III})~universal $m_0$, $\mhalf$, $A_0$ with NUHM relaxation and general $\alam$ and $\akap$. 

We use NMSSMTools-3.0.2~\cite{Ellwanger:2004xm}\cite{Ellwanger:2005dv}\cite{nmweb} 
for the numerical analysis, performing extensive scans over the parameter spaces of the models considered. The precise constraints imposed are the following.  Our `basic constraints' will be to require
 that an NMSSM parameter choice be such as to give a proper RGE solution, have no Landau pole, have a neutralino LSP and obey Higgs and SUSY mass limits as implemented 
in NMSSMTools-3.0.2 (Higgs mass limits are from LEP, TEVATRON, and early LHC data; SUSY mass limits are essentially from LEP). Regarding $B$ physics, the constraints considered are those on $\br(B_s\to X_s\gamma)$, $\Delta M_s$, $\Delta M_d$, $\br(B_s\to \mu^+\mu^-)$, $\br(B^+\to \tau^+\nu_\tau)$ 
and $\br(B\to X_s \mu^+ \mu^-)$ at $2\sigma$ as encoded in
NMSSMTools-3.0.2, except that we updated the bound on the radiative
$B_s$ decay to $3.04<\br(B_s\to X_s\gamma)\times 10^{4}<4.06$;
theoretical uncertainties in $B$-physics observables are taken into
account as implemented in NMSSMTools-3.0.2.  These combined
constraints we term the `$B$-physics contraints'.  Regarding $\amu$,
we require that the extra NMSSM contribution, $\delta \amu$, falls into 
the window defined in NMSSMTools of $8.77\times 10^{-10}<\delta\amu<
4.61\times 10^{-9}$ expanded to $5.77\times
10^{-10}<\delta\amu<4.91\times 10^{-9}$ after allowing for a $1\sigma$
theoretical error in the NMSSM calculation of $\pm 3\times
10^{-10}$. In fact, points that fail to fall into the above $\damu$ window always do so by virtue of $\damu$ being too small. For $\omghsq$, we declare that the relic density is
consistent with WMAP data provided $0.094<\omghsq<0.136$, which is the
`WMAP window' defined in NMSSMTools-3.0.2 after including 
theoretical and experimental systematic uncertainties.  We will also consider
the implications of relaxing this constraint to simply $\omghsq<0.136$
so as to allow for scenarios in which the relic density arises at
least in part from some other source. A ``perfect''  point will be one for which all constraints are satisfied including requring that $\delta \amu$ is in the above defined window and $\omghsq$ is in the WMAP window.

We find that only in models {\bf II} and {\bf III} is it possible for
a ``perfect'' point to have a light scalar Higgs in the mass range
$123-128\gev$ 
as consistent with the hints from the recent LHC Higgs searches. 
The largest $\mhi$ achieved for perfect points is about $125\gev$.  
However, relaxing the $\amu$ constraint vastly increases the number of
accepted points and it is possible to have  $\mhi\gsim 126\gev$  in
both models {\bf II} and {\bf III} even if $\damu$ is just slightly outside (below) the allowed window.  Comparing with \cite{Baer:2011ab}, the tension between obtaining an
ideal or nearly ideal $\damu$ while predicting a SM-like light Higgs
near 125 GeV appears to be somewhat less in NUHM variants of the NMSSM
than in those of the MSSM.

In the plots shown in the following, the coding for the plotted points is as follows:
\begin{itemize}
\itemsep=0in

\item grey squares pass the `basic' constraints but fail $B$-physics constraints (such points are rare);

\item green squares pass the basic constraints {\it and} satisfy $B$-physics constraints;

\item blue plusses ($+$) observe $B$-physics constraints as above and
  in addition have $\Omega h^2<0.136$, thereby allowing for other
  contributions to the dark matter density (a fraction of order
  20\% of these points have $0.094<\omghsq<0.136$) 
  but they do not necessarily have acceptable $\damu$;

\item magenta crosses ($\times$) have satisfactory $\damu$ as well as satisfying 
  $B$-physics constraints, but arbitrary $\omghsq$;

\item golden triangle points pass all the same constraints as the magenta points 
  and in addition have $\Omega h^2<0.136$;

\item open black/grey\footnote{For perfect points, we will use  black  triangles if $\mhi\geq 123\gev$ and grey triangles if 
    $\mhi < 123\gev$ in plots where $\mhi$ does not label the $x$ axis.} 
  triangles are perfect, completely allowed points in the sense that they pass all the constraints 
  listed earlier, including $5.77\times 10^{-10}<\damu<4.91\times 10^{-9}$ and $0.094<\omghsq<0.136$;

\item open white diamonds are points with $\mhi\geq 123\gev$ that pass basic constraints, $B$-physics
  constraints and predict $0.094<\omghsq<0.136$ but have $4.27\times
  10^{-10}<\damu<5.77\times 10^{-10}$, that is we allow an excursion of half the $1\sigma$ theoretical systematic uncertainty below the
  earlier defined window. We will call these ``almost perfect'' points.

\end{itemize}

The only Higgs production mechanism relevant for current LHC data is gluon-gluon to Higgs. For our plots it will thus be useful to employ the ratio of the $gg$ induced Higgs cross section times the Higgs branching ratio to a given final state, $X$, relative to the corresponding value for the SM Higgs boson:
\begin{equation}
R^{h_i}(X)\equiv {\Gamma(gg\to h_i) \ \br(h_i\to X)\over \Gamma(gg\to \hsm)\ \br(\hsm\to X)},
\end{equation} 
where $h_i$ is the $i^{th}$ NMSSM scalar Higgs, and $\hsm$ is the SM
Higgs boson. The ratio is computed in a self-consistent manner (that is,
treating radiative corrections for the SM Higgs boson in the same
manner as for the NMSSM Higgs bosons) using an
appropriate additional routine for the SM Higgs added to the NMHDECAY
component of the NMSSMTools package.   To compute the SM denominator,
we proceed as follows.\footnote{Ideally, the same radiative
corrections would be present in NMHDECAY as are present in
HDECAY~\cite{Djouadi:1997yw} and we could then employ HDECAY results
for the SM Higgs denominator.  But, this is not the case at
present, with HDECAY yielding, e.g., larger $gg$ production rates.  However, we note that since we compute the ratios of NMSSM rates to SM rates using the $C_Y$ couplings, as discussed below, the computed ratios will be quite insensitive to the precise radiative corrections employed.}   
NMHDECAY computes couplings for each $h_i$ defined by 
$C_Y^{h_i} \equiv g_{h_i  Y}/g_{\hsm Y}$, where $Y=gg,VV,b\bar
b,\tau^+\tau^-,\gam\gam,\ldots$, as well as $\Gamma_{\rm tot}^{h_i}$
and $\br(h_i\to Y)$ for all $Y$.  From these results we obtain the
partial widths $\Gamma^{h_i}(Y)=\Gamma_{\rm tot}^{h_i}\br(h_i\to Y)$.
We next compute $\Gamma^{\hsm}(Y)=\Gamma^{h_i}(Y)/[C_Y^{h_i}]^2$ and
$\Gamma_{\rm tot}^{\hsm}=\sum_Y \Gamma^{\hsm}(Y)$ and thence
$\br(\hsm\to Y)=\Gamma^{\hsm}(Y)/\Gamma_{\rm tot}^{\hsm}$.  We then
have all the information needed to compute $R^{h_i}$ for some given
final state $X$.

\begin{figure}
\vspace*{-.5in}
\centering
\includegraphics[width=0.74\textwidth]{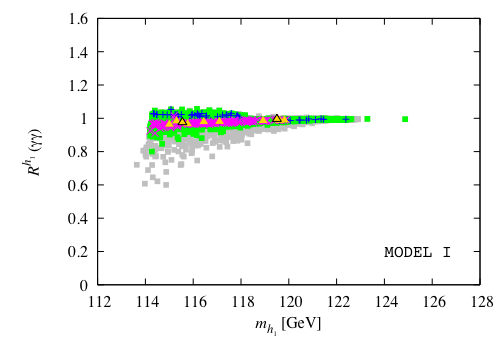}
\vskip -.15in
\includegraphics[width=0.74\textwidth]{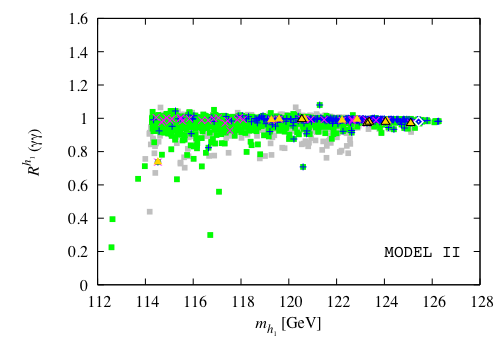}
\vskip -.15in
\includegraphics[width=0.74\textwidth]{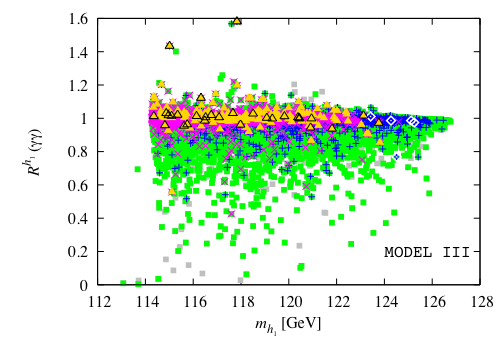}
\vspace*{-.1in}
\caption{Scatter plots of $\rhi(\gam\gam)$ versus $\mhi$ for boundary condition cases 
{\bf I}, {\bf II} and {\bf III}. See text for symbol/color notations. \label{rgamgamhi}}
\end{figure}

We begin by presenting the crucial plots of Fig.~\ref{rgamgamhi} in
which we show $\rhi(\gam\gam)$ as a function of $\mhi$ for cases 
{\bf I}, {\bf II} and {\bf III}.  
Only in cases {\bf II} and {\bf III} do we find points that pass all constraints 
(the open black triangles) with $\mhi\sim 124-125\gev$.  
These typically have $\rhi(\gam\gam)$ of order $0.98$. Somewhat
surprisingly, such points were more easily found by our scanning
procedure in case {\bf II} than in case {\bf III}. Many additional
points with $\mhi\sim 125\gev$ emerge if we relax only slightly the
$\damu$ constraint.  The white
diamonds show points for cases for which  $4.27\times 10^{-10}<\damu<5.77\times 10^{-10}$ 
having $\mhi\ge123\gev$.  As can be
seen in more detail from the sample point tables presented later, the
parameter choices that give the largest $\mhi$ values are ones for
which the $\hi$ is really very SM-like in terms of its couplings and
branching ratios.  Our scans did not find parameter choices for which
$\rhi(\gam\gam)$ was significantly larger than 1 for $\mhi=123-128\gev$, 
as hinted at by the ATLAS data.

As regards $\hii$, if we require $\mhii\in[110-150]\gev$ then we find
points that pass the basic constraints and the $B$-physics
constraints, but none that pass the further constraints. So, it
appears that within these models it is the $\hi$ that must be
identified with the Higgs observed at the LHC.
In contrast, if parameters are chosen at the SUSY scale without 
regard to GUT-scale unification, it is possible to find scenarios in 
which $\mhii\approx125\gev$ and, moreover, $\rhii(\gam\gam)>1$ \cite{Ellwanger:2011aa}.  

In passing, we note that should the Higgs hints disappear and a
low-mass SM-like Higgs be excluded then it is of interest to know if
$\br(\hi\to\ai\ai)$ can be large for $\mhi$ in the $\lsim 130\gev$
range. It turns out that, although large $\br(\hi\to\ai\ai)$ is possible
while satisfying basic and $B$-physics constraints, once additional constraints 
are imposed, $\br(\hi\to\ai\ai)\lsim 0.2$ for all three model cases being considered. 
Small $\br(\hi\to\ai\ai)$ is expected \cite{Dermisek:2006wr} (see also \cite{Dobrescu:2000yn}) 
when the $\ai$ is very singlet, as is the case in our scenarios once all constraints are imposed. 
So, in these models a light Higgs has nowhere to hide. 

The points in the scatter plots were primarily obtained through random scans
over the parameter spaces of the three models considered. In addition,
we performed Markov Chain Monte Carlo (MCMC) scans to zero in better
on points with $\mhi\sim 125\gev$ that observe all constraints.  For
this purpose, we defined a $\chi^2(m_{h_1})=(m_{h_1}-125)^2/(1.5)^2$.
The $B$-physics constraints were also implemented using a $\chi^2$ approach with the $1\sigma$ errors from theory and experiment (as implemented in NMSSMTools) combined in quadrature. 
 The global likelihood was then computed as $L_{\rm tot}=\prod_i L_i$ with $L_i=e^{-\chi_i^2/2}$ for two-sided constraints and 
$L_i=1/(1+e^{(x_i-x_i^{\rm exp})}/(0.01x_i^{\rm exp}))$ when $x_i^{\rm exp}$ is a 95\% CL upper limit. 
The $\amu$ and $\omghsq$ constraints were either implemented a-posteriori using the $2\sigma$ window approach of NMSSMTools, or also included in the global likelihood. Since CMSSM-like boundary conditions with $\alam=\akap=0$ did not generate points anywhere near the interesting region, we have only performed this kind of scan for cases {\bf II} and {\bf III}.  This allowed us to find additional ``perfect'' and ``almost perfect'' points for models {\bf II} and {\bf III} with $\mhi\gsim 123\gev$.

\begin{figure}
\centering
\includegraphics[width=0.9\textwidth]{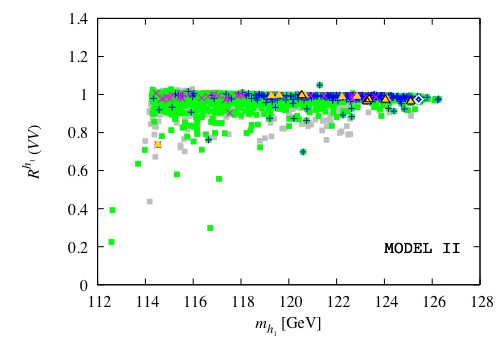}
\includegraphics[width=0.9\textwidth]{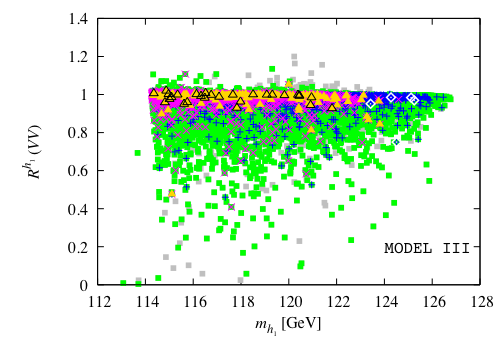}
\caption{Scatter plots of $\rhi(VV=ZZ,\,WW)$ versus $\mhi$ for models {\bf II} and  {\bf III}.  
See text for symbol/color notations. \label{zzww}}
\end{figure}

We next illustrate in Fig.~\ref{zzww} $\rhi(VV)$ (the ratio being the
same for $VV=WW$ and $VV=ZZ$) for boundary condition cases {\bf II}
and {\bf III}.  As for the $\gam\gam$ final state, for $\mhi\gsim 123\gev$
the predicted rates in the $VV$ channels are very nearly SM-like.
Overall, it is clear that, for the GUT scale boundary conditions
considered here, one finds that for parameter choices yielding
consistency with all constraints and yielding $\mhi$ close
to $125\gev$, the $\hi$ will be very SM-like.  If future data confirms
a $\gam\gam$ rate in excess of the SM prediction, then it will be
necessary to go beyond the constrained versions of the NMSSM
considered here (cf.\ \cite{Ellwanger:2011aa}).  And, certainly it 
is very difficult within the
constrained models considered here to obtain a SM-like Higgs with mass
much above $126\gev$ for parameter choices such that all constraints,
including $\damu$ and $\omghsq$, are satisfied.

\begin{figure}
\centering
\includegraphics[width=0.9\textwidth]{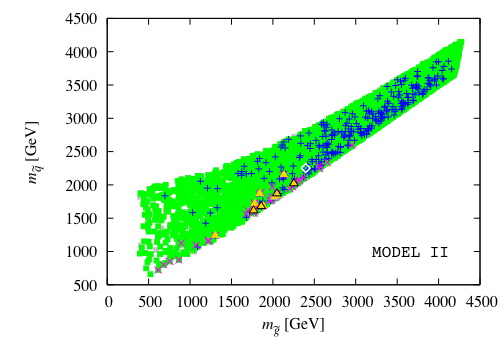}
\includegraphics[width=0.9\textwidth]{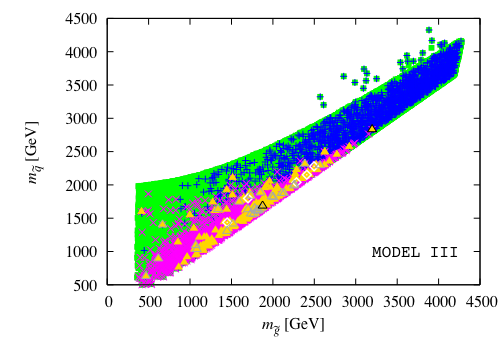}
\caption{Scatter plots of squark versus gluino masses for models {\bf II} and {\bf III}.  
Here we use black (grey) open triangles for perfect points with $\mhi \ge 123\gev$ ($\mhi < 123\gev$). 
See text for remaining symbol/color notations. \label{msqmgl}}
\end{figure}

Should a later LHC data set prove consistent with a rather SM-like Higgs in the vicinity of $\mhi\sim 125\gev$ (rather than one with an enhanced $\gam\gam$ rate), it will be of interest to know the nature of the parameter choices that yield the perfect, black triangle and almost perfect white diamond points with $\mhi \sim 125\gev$ and what the other experimental signatures of these points are.  We therefore present a brief summary of the most interesting features. 
First, one must ask if such points are consistent with current LHC limits on SUSY particles, in particular squarks and gluinos. To this end, 
Fig.~\ref{msqmgl} shows the distribution of squark and gluino masses for the various kinds of points for models {\bf II} and {\bf III}. Interestingly, all the perfect, black triangle and almost perfect, white diamond points with $\mhi\gsim 123\gev$ have squark and gluino masses above $1\tev$ and thus have not yet been probed by current LHC results. (Note that since we are considering models with universal $m_0$ and $\mhalf$ for squarks and gauginos, analyses in the context of the CMSSM apply.) 
It is quite intriguing that the regions of parameter space that are consistent with a Higgs of mass close to $125\gev$ automatically evade the current limits from LHC SUSY searches.

In order to further detail the
parameters and some relevant features of perfect and almost perfect points we present 
in Tables~\ref{goldtab1}--\ref{goldtab4} seven exemplary points with $\mhi\gsim 124\gev$ 
from models {\bf II} and {\bf III}. Some useful observations include the following:

\bit

\item Because of the way we initiated our model {\bf III} MCMC scans, restricting $|A_{\lambda,\kappa}|\le 1$~TeV,  most of the  tabulated model {\bf III} points have quite modest $\alam$ and $\akap$.  However, a completely random scan finds almost perfect points with quite large $\alam$ and $\akap$ values as exemplified by tabulated point \#7. The fact that the general scan over $\alam$ and $\akap$ did not find any perfect points with $\mhi\gsim 124\gev$, whereas such points were fairly quickly found using the MCMC technique, 
suggests that such points are quite fine-tuned in the general scan sense.  See Table~\ref{goldtab1} for specifics.

\item
In Table~\ref{goldtab2}, we display various details regarding the Higgs bosons for each of our exemplary points. As already noted, for the perfect and almost perfect points the $\hi$ is very SM-like when $\mhi\gsim 123\gev$.  To quantify how well the LHC Higgs data is described for each of our exemplary points, we use a chi-squared approach. In practice, only the ATLAS collaboration has presented the best fit values for $R^{h}(\gam\gam,ZZ\to4\ell,WW\to \ell\nu\ell\nu)$ along with $1\sigma$ upper and lower errors as a function of $\mh$. Identifying $h$ with the NMSSM $\hi$, we have employed Fig.~8 of \cite{atlashiggs} to compute a $\chi^2({\rm ATLAS})$ for each point in the NMSSM parameter space (but this was not included in the global likelihood used for our MCMC scans).  From Table~\ref{goldtab2} we see that the smallest $\chi^2({\rm ATLAS})$ values (of order 0.6 to 0.7) are obtained for $\mhi\sim 124\gev$. This is simply because at this mass the ATLAS fits to $R^{h}(\gam\gam)$ and $R^{h}(4\ell)$ are very close to one, the natural prediction in the NMSSM context.  For $\mh\sim 125\gev$, the $R^{h}$'s for the ATLAS data are somewhat larger than 1 leading to a discrepancy with the NMSSM SM-like prediction and a roughly doubling of $\chi^2({\rm ATLAS})$ to values of order 1.3 to 1.6 for our exemplary points.   In this context, we should note that at a Higgs mass of $125\gev$ the CMS data is best fit if the Higgs signals are not enhanced and, indeed, are very close to SM values.

\item The mass of the neutralino LSP, $\cnone$, is rather similar,  $m_{\cnone}\approx 300-450\gev$, for the different perfect and almost perfect  points with $\mhi\gsim 124\gev$. For all but pt.~\#5, the $\cnone$ is approximately an equal mixture of higgsino and bino. There is some variation in the  primary annihilation mechanism, with $\stau_1\stau_1$ and $\cnone\cnone$ annihilation being the dominant channels except for pt.~\#2 for which $\widetilde\nu_\tau\widetilde\nu_\tau$  and $\widetilde\nu_\tau\overline{\widetilde\nu}_{\tau}$ annihilations are dominant. In the case of dominant $\stau_1\stau_1$ annihilation, the bulk of the $\cnone$'s come from those $\stau$'s that have not annihilated against one another or co-annihilated with a $\cnone$. 

\item All the tabulated points yield a spin-independent direct detection cross section of order $(3.5-6)\times 10^{-8}~{\rm pb}$. For the above $\mcnone$ values, current limits on $\sigsi$ are not that far above this mark and upcoming probes of $\sigsi$ will definitely reach this level.

\item The 7 points all have $\mgl$ and $\msq$ above $1.5\tev$ and in some cases above $2\tev$.
Detection of the superparticles may have to await the LHC upgrade to $14\tev$.

\item Only the $\stopi$ is seen to have a mass distinctly below $1\tev$ for the tabulated points.
Still, for all the points $\mstopi$ is substantial,  ranging from $\sim 500\gev$ to above $1\tev$.  For such masses, detection of the $\stopi$ as an entity separate from the other squarks and the gluino will be quite difficult and again may require the $14\tev$ LHC upgrade.

\item The effective superpotential $\mu$-term, $\mu_{\rm eff}$, is small for all the exemplary points. 
This is interesting regarding the question of electroweak fine-tuning.   

\eit

\begin{table}[t]
\bec
\small{
\begin{tabular} {|c|c|c|c|c|c|c|c|}
\hline
&\multicolumn{3}{|c|}{Model II}&\multicolumn{4}{|c|}{Model III}\\\hline
Pt. \#      &   1*  &    2*  &   3  &    4*   &    5   &     6   &   7    \\ 
\hline \hline
$\tanb (m_Z)$&  17.9  &   17.8  &  21.4   &   15.1   &    26.2   &    17.9   &   24.2   \cr
$\lam$      & 0.078& 0.0096& 0.023 & 0.084  &  0.028  &  0.027  &  0.064 \cr
$\kap$      & 0.079& 0.011 & 0.037 & 0.158  & $-0.045$  &  0.020   &  0.343  \cr
$\mhalf$   & 923 & 1026  & 1087  &  842  &   738  &  1104    &  1143  \cr
$m_0$      & 447 & 297   & 809  &  244  &   1038  &  252   &  582  \cr
$A_0$       &$-1948$ & $-2236$  & $-2399$   & $-1755$  &  $-2447$  & $-2403$    & $-2306$   \cr
$\alam$    &  0   &  0    &  0    & $-251$  &  $-385$  & $-86.8$  & $-2910$  \cr
$\akap$    &  0   &  0    &  0    & $-920$  &   883  & $-199$   & $-5292$  \cr
$m_{H_d}^2$ & $(2942)^2$ & $(3365)^2$ & $(4361)^2$    &  $(2481)^2$  &   $(935)^2$  &  $(3202)^2$   & $(3253)^2$   \cr
$m_{H_u}^2$ & $(1774)^2$ & $(1922)^2$   & $(2089)^2$  &  $(1612)^2$   &  $(1998)^2$  &   $(2073)^2$   &  $(2127)^2$  \cr
\hline
\end{tabular}
}
\eec
\caption{Input parameters for the exemplary points. We give
  $\tanb(\mz)$ and GUT scale parameters, with masses in GeV and
  masses-squared in GeV$^2$.  Starred points are the perfect points satisfying all constraints, including $\damu>5.77\times 10^{-10}$ and $0.094<\omghsq<0.136$.  Unstarred points are the almost perfect points that have $4.27\times 10^{-10}<\damu<5.77\times 10^{-10}$ and $0.094<\omghsq<0.136$.  
\label{goldtab1}}
  \end{table}

\begin{table}[t]
\bec
\small{
\begin{tabular} {|l|c|c|c|c|c|c|c|}
\hline
&\multicolumn{3}{|c|}{Model II}&\multicolumn{4}{|c|}{Model III}\\\hline
Pt. \#                       &    1*   &    2*   &    3  &    4*   &    5   &    6   &   7   \\ \hline
 \hline
$\mhi$      &  124.0   &  125.1 &  125.4 &  123.8 &  124.5 &  125.2 &  125.1 \cr
$\mhii$     &  797   &  1011  &  1514  &  1089  &  430   &  663   &  302   \cr
$\mai$      &  66.5  &  9.83  &  3.07  &  1317  &  430   &  352   &  302   \cr
\hline \hline
$C_{u}$  & 0.999 & 0.999 & 0.999 &  0.999 &  0.999  & 0.999 &  0.999 \cr
$C_{d}$  & 1.002 & 1.002 & 1.001 & 1.003  &  1.139  & 1.002  & 1.002  \cr
$C_{V}$  & 0.999 & 0.999 & 0.999 & 0.999  &  0.999  & 0.999 &  0.999 \cr
$C_{\gam\gam}$  &1.003 & 1.004  & 1.004 & 1.004  & 1.012   & 1.003 & 1.001   \cr
$C_{gg}$  & 0.987 & 0.982 & 0.988 & 0.984  & 0.950   & 0.986 & 0.994
\cr
\hline
\hline
$\Gamma_{\rm tot}(h_1)$ [GeV]& 0.0037 & 0.0039 & 0.0039 & 0.0037 & 0.0046 & 0.0039 & 0.0039 \cr
$\br(h_1\to\gam\gam)$ 		  & 0.0024 & 0.0024 & 0.0024 & 0.0024 & 0.002  & 0.0024 & 0.0024 \cr
$\br (h_1\to gg)$ 			  &  0.056 &  0.055 &  0.056 &  0.056 & 0.043  & 0.055  & 0.056  \cr
$\br(h_1\to b\bar b)$ 		  &  0.638 &  0.622 &  0.616 &  0.643 & 0.680   & 0.619  & 0.621  \cr
$\br(h_1\to WW)$  			  &  0.184 &  0.201 &  0.207 &  0.180  & 0.159  & 0.203  & 0.201  \cr 
$\br(h_1\to ZZ)$			  & 0.0195 &  0.022 &  0.023 &
0.019 & 0.017  & 0.022  & 0.022  \cr
$R^{\hi}(\gam\gam)$ & 0.977 & 0.970 & 0.980 & 0.980 & 0.971 & 0.768 &
0.975 \cr
$R^{\hi}(ZZ,WW)$ & 0.971 & 0.962 & 0.974 & 0.974 & 0.964 & 0.750 &
0.969 \cr
\hline
$\chi^2_{\text{ATLAS}}$      &  0.59 &   1.27 &  1.47 & 0.72  & 1.57 &
1.34 & 1.20   \cr
\hline
\end{tabular}
}
\caption{Upper section: Higgs masses.  Middle section: reduced $h_1$ couplings to up- and down-type
  quarks, $V=W,Z$ bosons, photons, and gluons. Bottom section: total
  width in GeV, decay branching ratios, $R^{\hi}(\gam\gam)$,
  $R^{\hi}(VV)$ and $\chi^2_{\text{ATLAS}}$ of the lightest CP-even Higgs for the seven exemplary points.   \label{goldtab2}}
\eec
\end{table}

\begin{table}
\bec
\small{
\begin{tabular} {|c|c|c|c|c|c|c|c|}
\hline
&\multicolumn{3}{|c|}{Model II}&\multicolumn{4}{|c|}{Model III}\\\hline
Pt. \#      &   1*  &    2*  &   3  &    4*   &    5   &     6   &   7    \\ 
\hline \hline
$\mu_{\rm eff}$& 400 &  447   &  472   &  368   &  421   &  472   &
477   \cr
$\mgl$           & 2048 & 2253 & 2397 & 1876 & 1699 & 2410 & 2497 \cr 
$m_{\tilde q}$   & 1867 & 2020 & 2252 & 1685 & 1797 & 2151 & 2280 \cr
$m_{\tilde b_1}$ & 1462 & 1563 & 1715 & 1335 & 1217 & 1664 & 1754 \cr    
$m_{\tilde t_1}$ &  727 &  691 &  775 &  658 &  498 &  784 & 1018 \cr  
$m_{\tilde e_L}$ &  648 &  581 &  878 &  520 & 1716 &  653 &  856 \cr
$m_{\tilde e_R}$ &  771 &  785 & 1244 &  581 &  997 &  727 &  905 \cr
$m_{\tilde \tau_1}$&535 &  416 &  642 &  433 &  784 &  443 &  458 \cr 
$\mcpone$        &  398 &  446 &  472 &  364 &  408 &  471 &  478 \cr
$\mcnone$        &  363 &  410 &  438 &  328 &  307 &  440 &  452 \cr
\hline\hline
$f_{\tilde B}$  & 0.506 & 0.534 & 0.511 & 0.529  & 0.914  & 0.464  & 0.370  \cr
$f_{\tilde W}$  & 0.011 & 0.009& 0.008& 0.012  & 0.002 & 0.009 & 0.009 \cr
$f_{\tilde H}$  & 0.483 & 0.457 & 0.482 & 0.459  & 0.083  & 0.528  & 0.622 \cr
$f_{\tilde S}$  & $10^{-4}$ & $10^{-6}$  & $10^{-6}$ & $10^{-4}$ & $10^{-6}$ & $10^{-4}$ & $10^{-6}$ \cr
\hline
\end{tabular}
}
\caption{Top section: $\mu_{\rm eff}$ and
sparticle masses at the SUSY scale in GeV. Bottom section:  LSP decomposition. $m_{\tilde q}$ is the average squark mass of the first two generations. The LSP bino, wino, higgsino and singlino fractions are 
$f_{\tilde B}=N_{11}^2$,  $f_{\tilde W}=N_{12}^2$, 
$f_{\tilde H}=N_{13}^2+N_{14}^2$ and $f_{\tilde S}=N_{15}^2$, respectively, 
with $N$ the neutralino mixing matrix. 
\label{goldtab3}}
\eec
\end{table}

\begin{table}
\bec
\small{
\begin{tabular}{|c|c|c|c|c|c|}
\hline
Pt. \# & $\damu$ & $\Omega h^2$ & Prim. Ann. Channels & $\sigma_{\text{SI}}$ [pb] \\ \hline
1* & $6.01$ & 0.094 & $\cnone\cnone\to W^+W^- (31.5\%), ZZ (21.1\%)$ & $4.3\times 10^{-8}$ \cr 
2* & 5.85 & 0.099 & $\widetilde{\nu}_{\tau}\widetilde{\nu}_{\tau} \to \nu_{\tau}\nu_{\tau}(11.4\%), 
       \widetilde{\nu}_{\tau}\overline{\widetilde{\nu}}_\tau \to W^+W^- (8.8\%)$ & $3.8\times 10^{-8}$ \cr
3 & 4.48 & 0.114 & $\cnone\cnone\to W^+W^- (23.9\%), ZZ (17.1\%)$ & $3.7\times 10^{-8}$ \cr 
\hline
4* & 6.87 & 0.097 & $\cnone\cnone\to W^+W^- (36.9\%), ZZ (23.5\%)$ & $4.5\times 10^{-8}$ \cr 
5  & 5.31 & 0.135 & $\cnone\cnone\to b \bar{b}(39.5\%), h_1 a_1 (20.3\%)$ & $5.8\times 10^{-8}$ \cr  
6  & 4.89 & 0.128  & $\stau_1\stau_1\to \tau\tau(17.4\%),\cnone\cnone\to W^+W^- (14.8\%)$&$4.0\times10^{-8}$ \cr
7  & 4.96 & 0.101 & $\cnone\cnone\to W^+W^- (17.7\%), ZZ (12.9\%)$ & $4.0\times 10^{-8}$ \cr   
\hline
\end{tabular}
}
\caption{$\damu$ in units of $10^{-10}$, LSP relic abundance, primary
  annihilation channels and spin-independent LSP scattering cross section off protons. \label{goldtab4}}
\eec
\end{table}

For completeness, we have run separate scans for the case of the cNMSSM of~\cite{Djouadi:2008yj}\cite{Djouadi:2008uj} with completely universal $m_0=0$ and $A_0\equiv A_t=A_b=A_\tau=A_\lambda=A_\kappa$ (which is in fact a limit case of our model {\bf III}). 
Here, one can have a singlino LSP. This requires small $\lambda<10^{-2}$. Correct relic density is achieved via co-annihilation with $\tilde \tau_R$ for the rather definite choice of $A_0\sim -\frac{1}{4} M_{1/2}$. 
For small enough $\mhalf$, the $\hi$ is dominantly singlet, while the $\hii$ is SM-like. For larger $\mhalf$, the $\hi$ is SM-like, and the $\hii$ is mostly singlet. The cross-over where $\hi$ and $\hii$ are highly mixed occurs roughly in the range of $\mhalf=500-600\gev$, depending on $\lambda$. 
Overall, we find that the $\hi$ can attain a mass of at most $\sim 121$~GeV in this scenario in the limit of large $\mhalf$.\footnote{A similar conclusion was reached in \cite{Arbey:2011ab}  based on a mSUGRA scenario  with $m_0\approx 0$ and $A_0\approx -\frac{1}{4} M_{1/2}$, which  approximately corresponds to the  cNMSSM case with the singlet Higgs superfield decoupling from the rest of the spectrum; a maximum $h^0$ mass of $123.5$~GeV was found in this case.}
The $\hii$, on the other hand, can have a mass in the $123-128\gev$ range for not too large $\mhalf$. 
For $\lambda=10^{-2}$, this happens in the region of the cross-over where $\rhii(\gam\gam)$ is of order $ 0.5-0.6$. 
Squark and gluino masses are around $1.2-1.3$~TeV in this case, and hence highly pressed by LHC exclusion limits.  For smaller $\lambda$, an $\hii$ with mass near $125\gev$ is always singlet-like and its signal strength in the $\gamma\gamma$ and $VV$  channels is very much suppressed relative to the prediction for the SM Higgs, in apparent contradiction to the ATLAS and CMS results. \\

{\bf In summary}, we find that the fully constrained 
version of the NMSSM is not able to yield a Higgs boson consistent with the current hints from LHC data for a fairly SM-like Higgs with mass $\sim 125\gev$, once all experimental constraints are imposed including acceptable $\amu$ and $\omghsq$ in the WMAP window.  However, by relaxing the CNMSSM to allow for non-universal Higgs soft-masses-squared (NUHM scenarios), it is possible to obtain quite perfect points in parameter space  satisfying all constraints with $\mhi\sim 125\gev$ even if the attractive $U(1)_R$ symmetry limit of $\alam=\akap=0$ is imposed at the GUT scale and certainly if general $\alam$ and $\akap$ values are allowed. We observe a mild tension between 
the $\amu$  constraint and obtaining $\mhi\sim 125 \gev$; just slightly relaxing the $\amu$ 
requirement makes it much easier to find viable points with $\mhi\sim 125\gev$, thus opening up interesting regions of parameter space. 
We also note that our scanning suggests that relatively small $\alam,\akap$ values are preferred for (almost) perfect points. Masses of SUSY particles for perfect/almost perfect points are such that direct detection of SUSY may have to await the $14\tev$ upgrade of the LHC.  However, the predicted $\cnone$ masses and associated spin-independent cross sections suggest that direct detection of the $\cnone$ will be possible with the next round of upgrades to the direct detection experiments.

\bigskip
\noindent{\bf Acknowledgements:} 
We would like to thank Sezen Sekmen for helpful contributions regarding the MCMC program structure. 
This work has been supported in part by US DOE grant DE-FG03-91ER40674 and by  
IN2P3 under contract PICS FR--USA No.~5872.



\end{document}